\documentstyle[11pt]{article}
\begin{document}
\title{\huge The  Effect Of Delay Times On The Optimal Velocity Traffic Flow Behavior} 
\author{\bf H. Ez-Zahraouy$^1$, Z. Benrihane, A. Benyoussef\\  
\it Laboratoire de Magn\'{e}tisme et de la Physique des Hautes Energies,\\
\it Departement de physique, Facult\'e des Sciences, B.P.1014, Rabat, Morocco\\
}
\date{ }
\maketitle
\begin{flushleft}                
\large\bf{Abstract}
\end{flushleft}
We have numerically investigated the effect of the delay times $\tau_f$ and $\tau_s$ of a mixture of fast and slow vehicles on the fundamental diagram of the optimal velocity model. The optimal velocity function of the fast cars depends not only on the headway of each car but also on the headway of the immediately preceding one. It is found that the small delay times have almost no effects, while, for sufficiently large delay time $\tau_s$ the current profile displays qualitatively five different forms depending on $\tau_f$, $\tau_s$ and the fractions $d_f$ and $d_s$ of the fast and slow cars respectively. The velocity (current) exhibits first order transitions at low and/or high densities, from freely moving phase to the congested state, and from congested state to the jamming one respectively accompanied by the existence of a local minimal current. Furthermore, there exist a critical value of $\tau_f$ above which the metastability and hysteresis  appear. The spatial-temporal traffic patterns present more complex structure. 

\renewcommand{\thefootnote}{}
\footnote{} \renewcommand{\thefootnote}{\arabic{footnote}}
\footnotetext[1]{corresponding author E-mail address: ezahamid@fsr.ac.ma} 

\newpage

\section{\protect\bigskip Introduction}

Traffic flow problem has been extensively studied from physical point of view. Fluid-dynamical model [1], cellular automaton model [2], and car following model [3] have been proposed and analyzed in details to understand the mechanism of the traffic congestion on a freeway, using both analytical approaches and numerical simulations based on concepts and techniques of statistical physics [4,5,6] in one and two dimensions, in order to make clear the characteristics of this complex behavior [7,8].

Traffic flow phenomena strongly depend on the occupancy of the road. The nature of the variation of the flow with the density is still not clearly understood [9] since the details of the complex experimental set-up can strongly influence the empirical results [10,11].
In order to use the empirical results for a theoretical analysis it is often more convenient to use the mean values of the flow at a given density, a collection of possible forms at averaged fundamental diagrams consistent with empirical data [9], while the discontinuity of the fundamental diagram now seems to be well established [12] no clear answer can be given to the question on the form of the diagram in the free flow or high density regime. In the low density regime, linear as well as non linear functional forms of the fundamental diagrams have been suggested, for the high density branch no consistent picture for the high density branch exists. Here the results strongly depend on the specific road network.

Toward a realistic model which explains the traffic flow dynamics, the optimal velocity (OV) model proposed by Bando et al. [13,14] has attracted considerable interest. Based on the second-order differential equations, the model reveals the density pattern formation of the congested flow of traffic without introducing a time lag caused by the driver's response. Although the OV model is shown to have a universal structure in spatial-temporal patterns in the congestion, most of the analyses of the model have been done in the case where the optimal velocity function depends only on the headway of each car. One of the approaches to generalize the OV model is that the backward reference function is introduced [15]. Another approach to extend the OV model is to take into account the next-nearest-neighbor interaction [16-21], where the optimal velocity function depends not only on the headway of each car but also on the immediately preceding one. The generalized optimal velocity function is determined by taking into account the driver's skill, experience, and psychological effect, so that it is expected to describe more realistic traffic flow.

In a previous works we have study the effect of open boundaries in the optimal velocity model with delay time [22] in which we have established the phases diagrams in the injecting and extracting rate plane. Our aim in this paper is to investigate, using numerical simulations with parallel dynamics, the effect of the delay times $\tau_f$ and $\tau_s$ of the mixture of fast and slow vehicles on the fundamental diagram of the optimal velocity model [16,17] in the one dimension. Similar works have been proposed using cellular automaton model [23] to illustrate the same idea in order to give a clear explanation of the empirical data for flow and density [9,10,11].
The paper is organized as follows; in the following section the historic of the optimal velocity model is reviewed and we define our contribution, the section 3 is reserved for results and discussions, the conclusion is given in section 4.

\section{The Model}

We consider a one-dimensional road of length L with periodic boundaries, we can express the driving strategy of the driver in the car following models in terms of mathematical symbols by writing
\begin{equation}
\ddot x_{n}(t)=\frac{1}{\tau}\lbrack V^{desired}_{n}(t)- v_n(t) \rbrack 
\end{equation}
where $V^{desired}_{n}(t)$ is the desired speed of the n-th driver et time t. in all follow-the-leader models the driver maintains a safe distance from the leading vehicle by choosing the speed of the leading vehicle as his/her own desired speed in the case in the absence of overtaking i.e., $V^{desired}_{n}(t)=\dot x_{n+1}$. An alternative possibility has been explored in recent works based on the car-following approach [18,19]. This formulation is based on the assumption that $V^{desired}$ depends on the distance-headway of the n-th vehicle, i.e., $V^{desired}_{n}(t)=V^{opt}(t)$ so that
\begin{equation}
\ddot x_{n}(t)=\frac{1}{\tau}\lbrack V^{opt}(\Delta x_{n}(t))- v_n(t) \rbrack 
\end{equation}
Where the so-called optimal velocity function $ V^{opt}(\Delta x_{n}(t))$ depends on the corresponding instantaneous distance-headway $\Delta x_n (t)=x_{j+1}(t)-x_j(t)$, where $x_n(t)$ is the position of the vehicle n at time t. In order words, according to this alternative driving strategy, the n-th vehicle tends to maintain a safe speed that depends on the relative position, rather than relative velocity, of the n-th vehicle. In general, $V^{opt}(\Delta x_{n})\to 0$ as $ \Delta x \to 0 $ and must be bounded for $\Delta x \to \infty$. The optimal velocity function have been proposed by Bando et al. [13] is given by 
\begin{equation}
 V^{opt}(\Delta x_n (t))=tanh(\Delta x_n (t)-h_c)+tanh(h_c) 
\end{equation}
Where $h_c = 5$ is the safety distance 
Furthermore, by transforming the time derivative to the difference in Eq (2), one can obtain the difference equation  model .\\
\begin{equation}
x_n(t+2 \tau)=x_n(t+\tau)+\tau V^{opt}(\Delta x_n (t))  
\end{equation}
In order to study the traffic consisting of two different types of vehicles, say, cars and trucks, Mason et al. [20] generalized the formulation of Bando et al. [13] by replacing the constant $\tau$ by $\tau_n$, with $n=f$ or $n=s$ on whether the n-th vehicle is fast or slow as for example the case of  cars or trucks. Since a truck is expected to take longer to respond than a car, we should assign larger $\tau_s$ to trucks and smaller $\tau_f$ to cars. Some other mathematically motivated generalizations of the OV model have also been considered [18,21].

On the road the vehicles run with different responses and reactions, we consider $d_f$ the fraction of fast vehicles with smaller delay time $\tau_f $, while the remaining fraction $d_s$ is assigned to slow vehicles, i.e., vehicles with larger delay time $\tau_s $, where $\rho_T$ is the global density. The situation of overtaking is considered only when the following criteria are satisfied:  the driver sometimes pays attention to not only the headway but also the headway of the immediately preceding one, hence, if the fast vehicle follows the slow one the OV function take into account the headway of the next-nearest-neighbor interaction, than the OV function of the fast vehicle is given by
$$V^{opt}_{f\to s}(\Delta x_f (t),\Delta x_s(t))=tanh(\Delta x_n (t)+\Delta x_{n+1}(t)-h_c)+tanh(h_c)\eqno(3)$$

\section{Results and Discussion}

In our simulation the rule described above is updated in parallel dynamics, i.e. during one updating procedure the new particle positions do not influence the rest and only previous positions have to be taken into account. A homogeneous distribution of standing vehicles is used as the initial condition while the headway  are chosen as follows: $ \Delta x_j(0) = \Delta x_0$ for all j, where the initial number of particles is $N= \rho_T L$. Cars are numbered as 1,2,3,....N. In order to equilibrate the system, 50000 Monte-Carlo steps have been performed to ensure that steady state is reached, then, the average current and velocity are collected.

In the following we discuss the impact of the delay time of slow vehicles, for a fixed fraction and delay time of fast ones on the behavior of current and velocity as a function of density. However, it is found that the fundamental diagram exhibits and five different kind of topologies, depending on the values of $\tau_s$ namely:\\
i)For $\tau_s<\tau_s1$, the current (velocity) increases with density, passes through a maximum at a critical density, decreases smoothly and vanishes continuously. This case is presented in Fig.1a (Fig.1b), but for $\tau_s$ smaller than 0.8. In this case we note that the fundamental diagram exhibits a second order transition from free moving phase to the congested one since the tangent at the maximal current is non null, then at this point the velocity and current have two different "half tangents" [24]\\
ii)For $\tau_s1<\tau_s<\tau_s2$, the current (velocity) exhibits one first order transition which is characterized by the discontinuity of the current and velocity [25-27] at a critical density from the congested phase to the jamming one. Such behavior is shown in Fig.1a ( Fig.1b) for $0.8<\tau_s<1.6$ \\
iii)For $\tau_s2<\tau_s<\tau_s3$, the current and velocity display two first order transitions accompanied by a maximal and minimal value of the current at this transition (see Fig.2a). The first transition is located at low density, from freely moving phase to the congested one, while the second is located at high density, from congested to the jamming one.\\
 iv)For $\tau_s3<\tau_s$, the current exhibits only one first order transition at low density, from freely moving phase to the congested one. The current passes through a local maximum and a local minimum at the transition point. The absolute maximum of the current is located at congested phases (intermediaries densities) but the later is not a phase transition, it is only a crossover, since the tangent at the maximum is zero. This situation is illustrated in Fig.2b\\ 
v)For $\tau_s>\tau_s3$, the transition from freely moving phase to the congested one is still exist, accompanied by a small jump of the velocity and current at the transition point (Fig.2c), but the local minimal current disappears.\\
Furthermore, Fig.1 shows that the maximal current increases when increasing the value of the delay time $\tau_s$, this is due to the increase of the probability allowed to the fast vehicle to overtake the slow one, while for small value of $\tau_s$, this probability becomes very small, which lead to the blockage of fast vehicle, then the maximal current decreases.  The values of $\tau_s1$, $\tau_s2$ and $\tau_s3$ depend on the fraction of slow $d_s$ and fast $d_f$ vehicles and/or on the delay time $\tau_f$. Indeed, for $\tau_f = 0.5 $, $d_f=0.2$, the fundamental diagram exhibits the topology (ii) for any value of $\tau_s$ smaller than $1.6$, while for greater values of $\tau_s$ (Fig.2a), the fundamental diagram exhibits the topology (iii).\\
Beside in Fig.1a  shows also the existence of two different branches at free flow, one with positive slope (increasing current) and the other with negative one (decreasing current) which meet at local maximal current. At congested traffic flow, obviously completely blocked states exist for densities $\rho_T > 0.5$, therefore the flow in the stationary state is zero. When increasing the delay time of the slow cars the negative branch becomes a platoon formation where the flow $<J>(\rho_T$) is independent of the density and same thing about $<V>(\rho_T)$ (see Fig.1b), that is similar to the presence of a defect in the system due to the strong effect of the slow cars, such situation leads to the formation of coherent moving blocks of vehicles led by a slow cars. 

For high values of the delay time $\tau_s > 1.6$. Fig.2a-2c illustrate the effect of the fraction of fast vehicles with $\tau_s = 2.5$ and $\tau_f = 0.5$, Indeed, the fundamental diagram exhibits in the free flow density regime two branches of current with positive slopes and a platoon current for intermediate densities. Moreover, we can distinguish between several different behaviors, depending on the value of the density $\rho_T$ of vehicles in the road namely: \\
For $\rho_T \leq \rho_{max1}$ , the current increases with $\rho_T$ to reach at $\rho_{max1}$ its first maximal value $I_{max1}$ which depend strongly on the values of $d_f$ , indeed, in this region the overtaking rule is satisfied due to lower values of $\rho_T$.\\
For $\rho_{max1} \leq \rho_{min}$; $\rho_{min}$ is the density at which the current present a nonzero minimum $I_{min}$. In this region (the congested phase of fast cars), the situation of 
overtaking becomes more difficult as long as $\rho_T$ increases, i.e, not all fast vehicles can overtake, hence the current value fall with $\rho_T$ which means a competition between 
platoon phenomena and overtaking. Note that $I_{min}$ does not depend on $d_f$. This is due to the structure of the chain in this region, i.e, there is no sufficiently empty space for overtaking in comparison to the first region. Indeed, for $\rho_T \geq \rho_{min}$ (the third region) fast cars can not overtakes slow cars, such situation leads to the formation of coherent moving blocks of vehicles each of which is led by slow cars, i.e, the phenomena of platoon formation. We conclude that the first maximum $I_{max1}$ corresponds to the transition from free flow to congested phase of type1. $I_{min}$ indicates the end of the competition .

In order to establish the existence of metastable states and hysteresis effects one can follow two basic strategies. In the first method, the density of cars is changed adiabatically by adding or removing vehicles from the stationary state at a certain density. starting in the jamming phase (large densities) and removing cars, one obtains the lower branch of the hysteresis curve see Fig.3. On the other hand, by adding cars to a free flowing state (low densities), one obtains the upper branch. The second method, does not require changing the density. instead one starts from two different initial conditions, the mega-jam and the homogeneous state. The mega-jam consists of one large, compact cluster of standing cars. In the homogeneous state, cars are distributed equidistantly ( with one large headway for incommensurable densities). In certain density regimes the FD can consist of two branches. in the upper branch ( with higher flow) there are almost no interactions between the cars and the system in a homogeneous, jam-free state. In the lower branch, however, the system is in a "phase-separated" state, consisting of one large jam and a free flowing part.

To illustrate the first order transition from the congested state to the jamming one obtained in Fig.1a, for $\tau_f =0.5$, $\tau_s =1.59$, and $d_f = 0.8$, we have plotted  the space-time evolution traffic patterns, in Figs.4 for different values of the global density $\rho_T$, in which three different kind of  structure appear, depending on the value of the global density. Indeed, Fig. 4a shows that the local jam begin to persist,  but with a small fraction of vehicle. While, near the congested-jamming transition ($\rho_T=0.58$), the local jam becomes more important and the  size of the cluster of the jam increases, but the average velocity of the moving vehicle is still important (Fig.4b), such situation is interpreted as the coexistence between jam and moving phases. For $\rho_T=0.61$ the jamming phase take place and the average velocity of vehicle vanishes (Fig.4c).\\

\section{Conclusion}
In summary, we have studied using numerical simulations the effects of the mixture of two kinds of vehicles with different delay times of driver's response $\tau_f$ and $\tau_s$ in which fast vehicle can overtakes the slow one. Depending on the values $\tau_s$, it is found that the fundamental diagram exhibits first order and/or second order transitions from freely moving phase to the congested one and from congested phase to the jamming one. The fundamental diagram present also the minimal current for intermediate values of $\tau_s$ and disappears otherwise. Beside, the model exhibits the metastability and hysteresis phenomena under certain conditions.    

\bf  ACKNOWLEDGMENT \\ 
\rm  This work was financially supported by the Protars II n° P11/02.\\

\newpage \textit{Figure captions:}\\

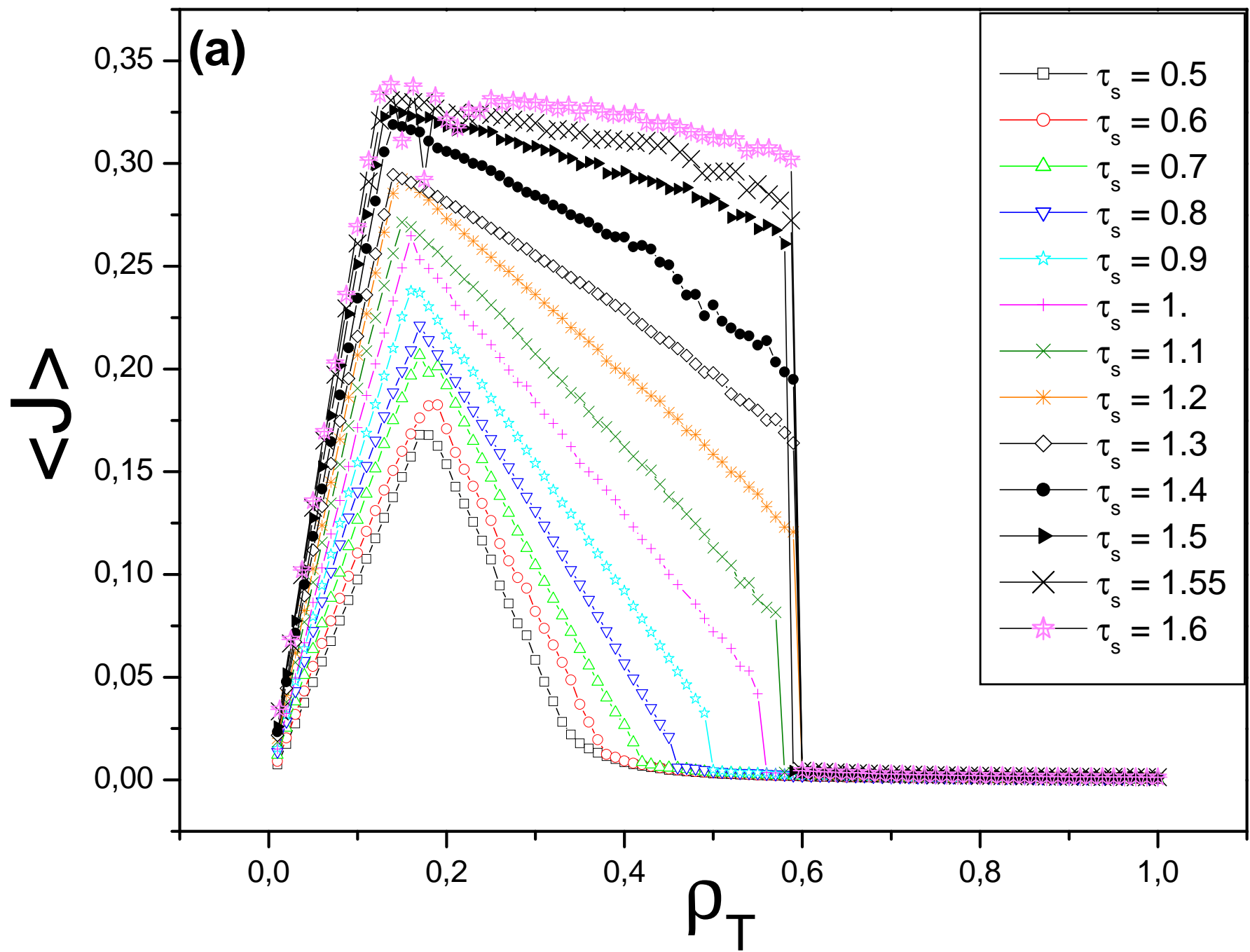
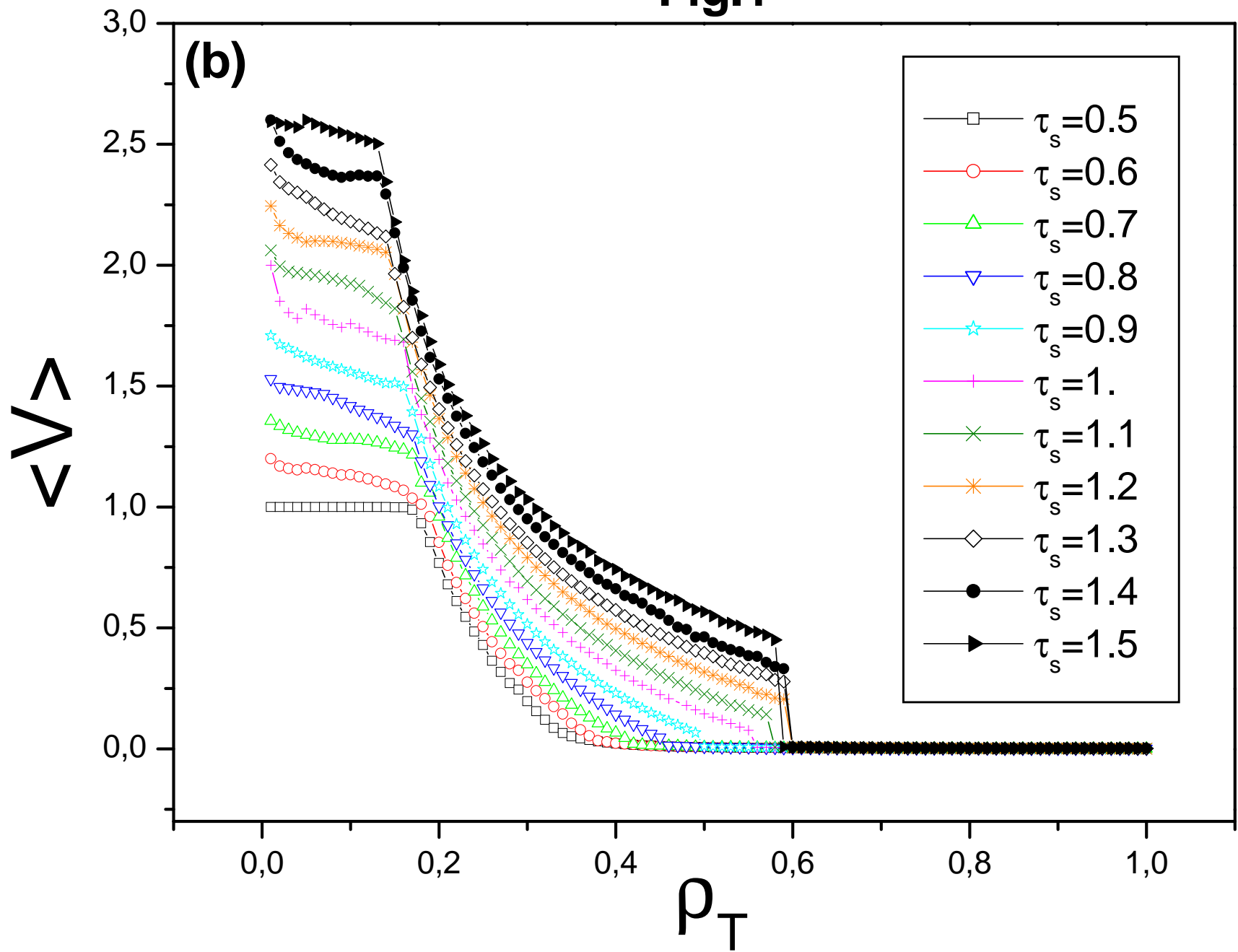
Fig.1 : The mean current and the average velocity as a function of car density in (a) and (b) respectively, for two types of vehicles with $d_f = 0.2$, $\tau_ f =0.5 $ and
for several values of delay time $\tau_{s} $  

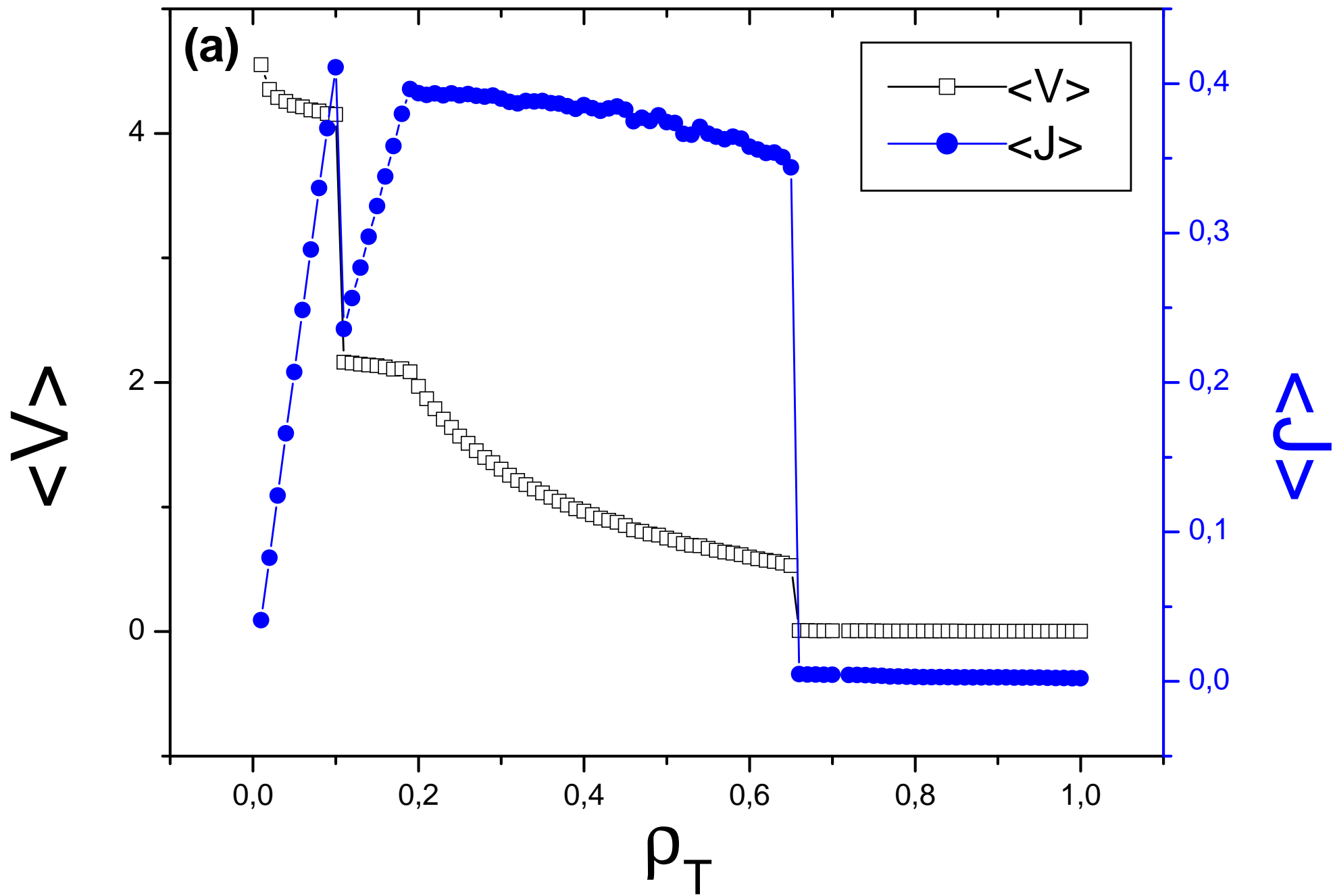
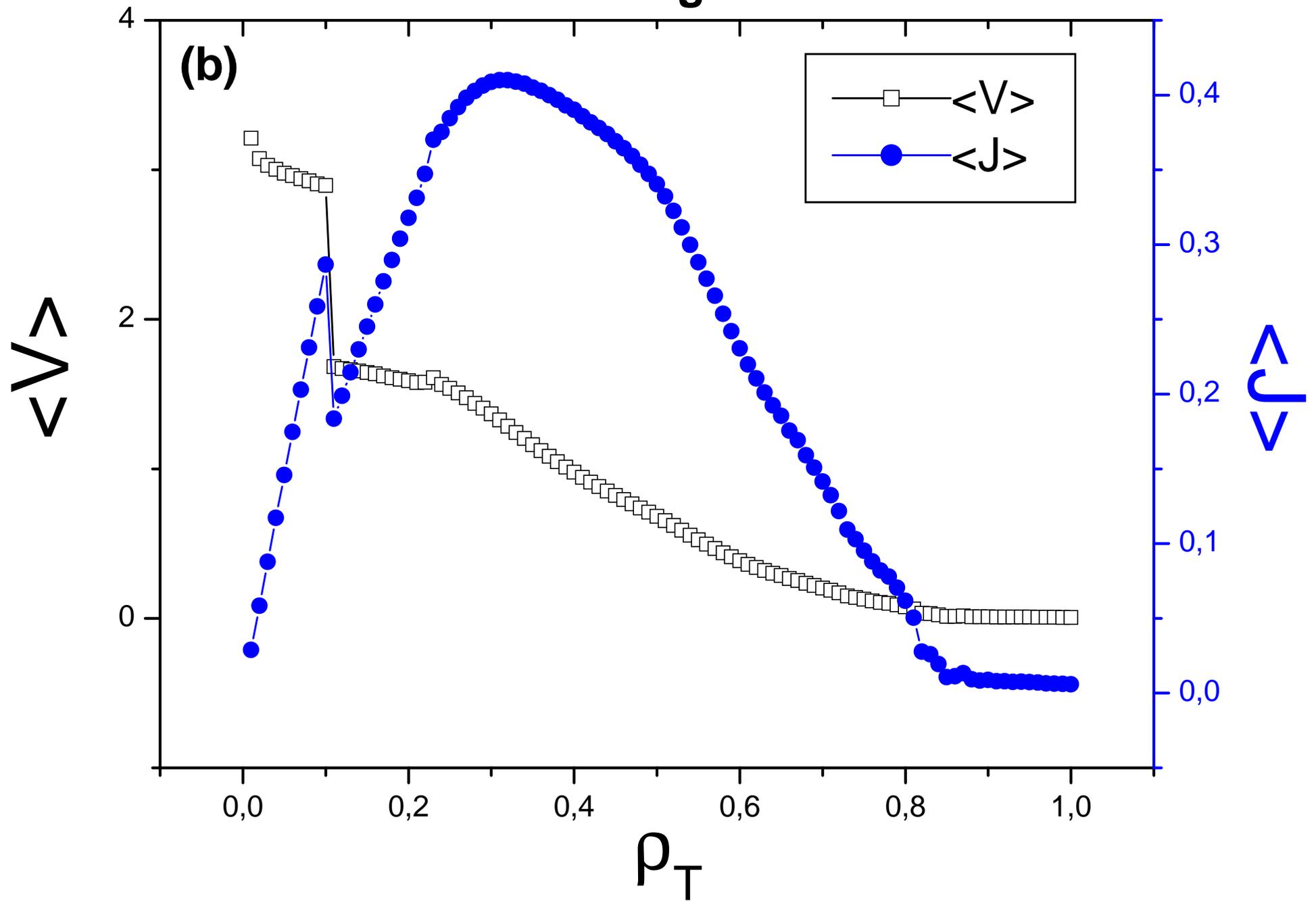
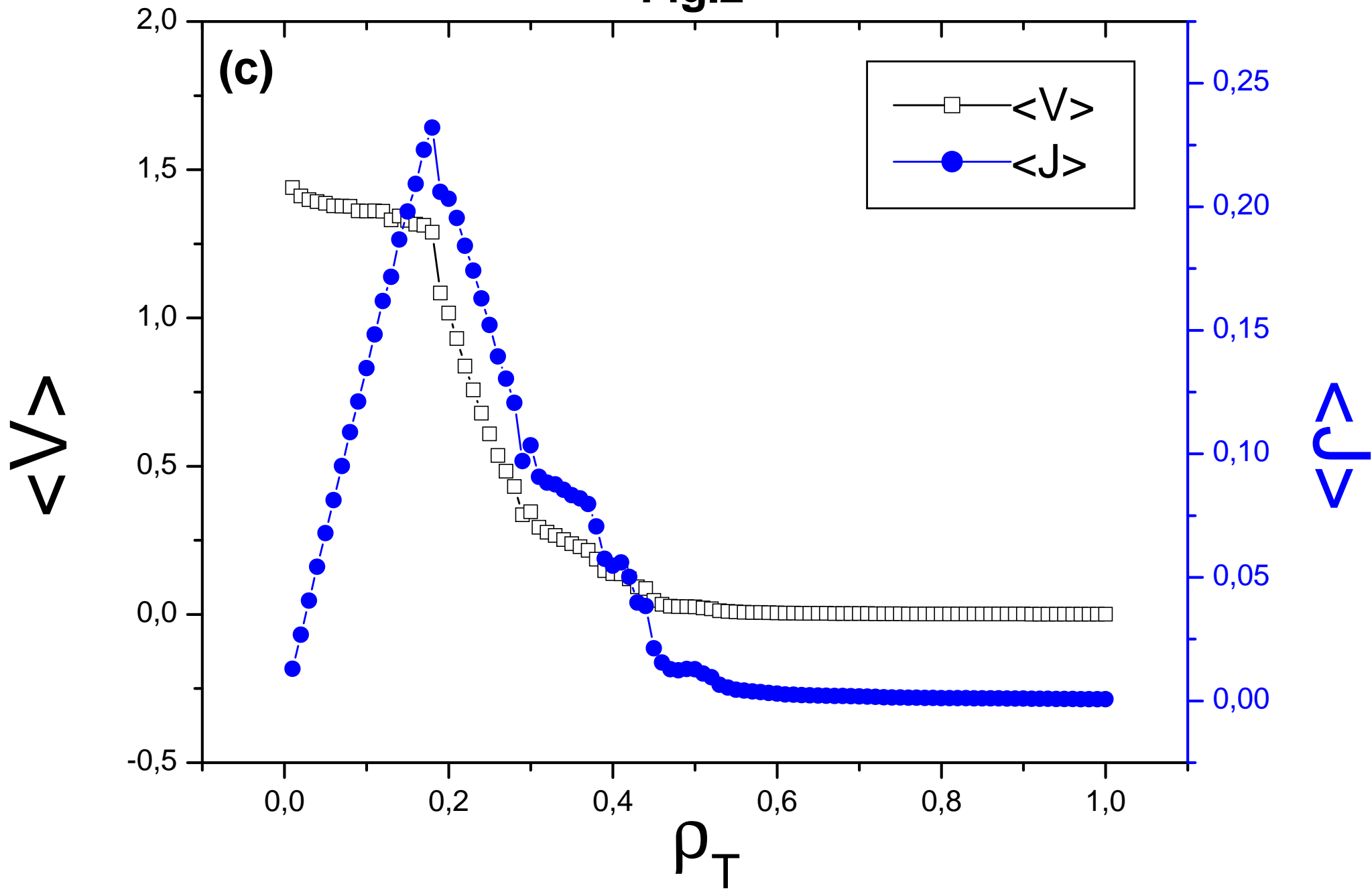
Fig.2 : The fundamental diagram with the average velocity versus the density for $\tau_f =0.5$ and $\tau_s =2.5$; (a) $d_f = 0.2$; (b) $d_f = 0.5$ and (c)$d_f = 0.8$ 

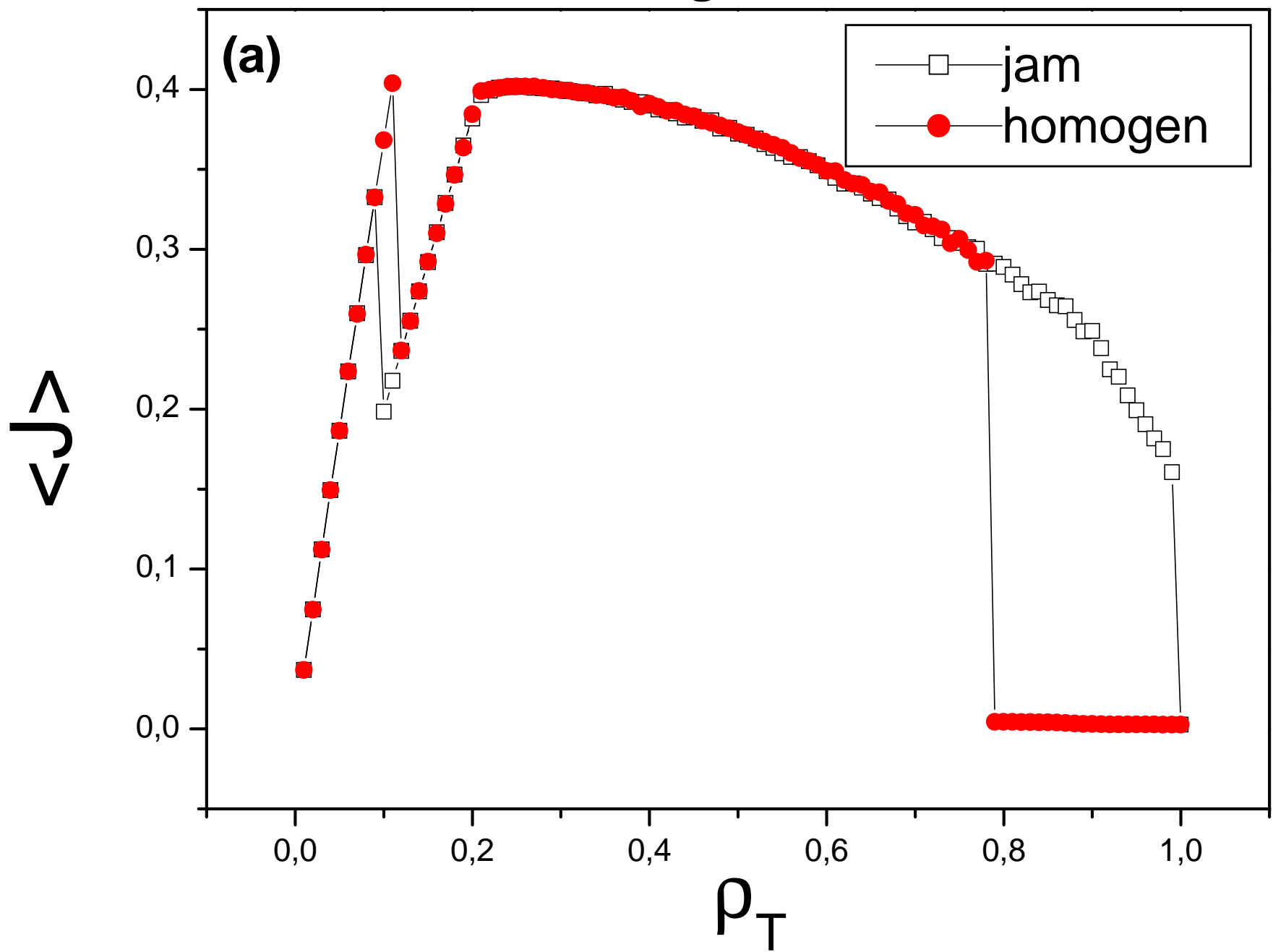
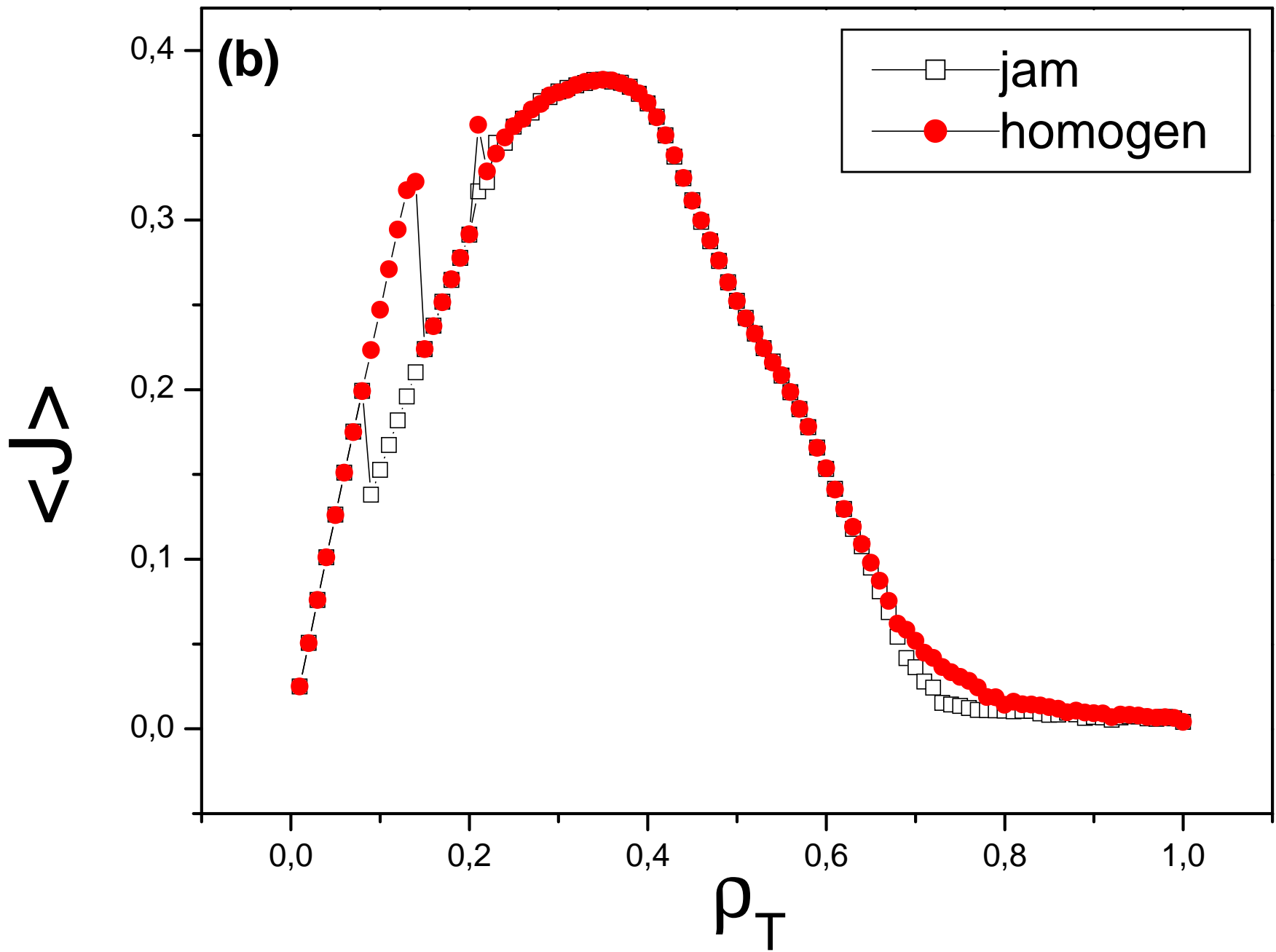
Fig.3 : The fundamental diagram for homogeneous (circle dots) and jam (square dots) initial configurations for $\tau_f =0.5$  and $\tau_s =2.5$; (a) $d_f = 0.3$; (b) $d_f = 0.6$

Fig.4 : The space-time evolution for, $\tau_f =0.5$, $\tau_s =1.59$ and $d_s = 0.2$,  (a) $\rho_T = 0.18$, (b) $\rho_T = 0.58$, (c) $\rho_T = 0.61$. The solid lines present the trajectory of slow cars and the dots correspond to the fast ones, the dark region exhibits the jam.

\end{document}